# ADVANCES TOWARDS PB11 FUSION WITH THE DENSE PLASMA FOCUS

Eric J. Lerner
Lawrenceville Plasma Physics
Robert E. Terry
Naval Research Laboratory

**ABSTRACT**

The prospects for achieving net energy production with pB11 fuel have recently considerably brightened. Studies have shown that the multi-GG field potentially obtainable with modest dense plasma focus devices have the effect of reducing the flow of energy from the ions to the electrons and thus suppressing bremsstrahlung radiation that cools the plasma[1]. We report here on new simulations that indicate that net energy production may be achievable in high-magnetic-field devices at peak currents as low as 2.3 MA. While these simulations only model the dense plasmoid formed in the focus, new simulation techniques can allow a full particle-in-cell simulation of DPF functioning over the wide range of time and space scales needed. Such simulations will be of great value in the next round of experiments that will use pB11 fuel.

1. **INTRODUCTION**

Controlled fusion with advanced fuels, especially hydrogen-boron-11, is an extremely attractive potential energy source. Hydrogen-boron fuel generates nearly all its energy in the form of charged particles, not neutrons, thus minimizing or eliminating induced radioactivity. The main reaction, $p+^{11}B \rightarrow 3\,^{4}He$, produces only charged particles. A secondary reaction, $^{4}He+^{11}B \rightarrow\ ^{14}N + n$ does produce some neutrons as the alpha particles produced by the main reaction slow down in the plasma, but only about 0.2% of the total fusion energy is carried by the neutrons, whose typical energy is only 2.5 MeV. Hydrogen-boron fuel also allows direct conversion of charged-particle energy to electric power, without the expensive intermediate step of generating steam for turbines.[1-3] While this fuel requires extremely high ion energies, above 200 keV, there is evidence that such energies can be achieved in the dense plasma focus [1] as well as in the z-pinch [4].

However, because of the $z^2$ dependence and boron's z of 5, bremsstrahlung x-ray radiation is enhanced for $p^{11}B$ fuel. Many analyses have indicated that fusion power can barely if at all exceed plasma cooling by bremsstrahlung [5]. If unavoidable, this situation would eliminate the heating of the plasma by the fusion-produced alpha particles and would require that all the energy be recovered from the x-ray radiation.

But published analyses have overlooked an important physical effect that is especially relevant for the use of $p^{11}B$ with the dense plasma focus. This effect, first pointed out by McNally [6], involves the reduction of energy transfer from the ions to the electrons in the presence of a strong magnetic field. This in turn reduces the electron temperature and thus the bremsstrahlung emission.

For ions colliding with electrons with gyrofrequency $\omega_g$, energy transfer drops rapidly for impact parameters $b > v_i/\omega_g$, where $v_i$ is ion velocity, since in that case the electron is accelerated at some times during the collision and decelerated at others, so average energy transfer is small. This means that the $b_{max}$ is less than the Debye length, $\lambda_D$ by a factor of $v_i\,\omega$ $_p$/ $v_{et}$ $\omega_g$, where $\omega_p$ is the plasma frequency and $v_{et}$ is the electron thermal velocity. So the Coulomb logarithm in the standard energy-loss formula is reduced to Ln ($mv_i^2/\hbar\omega_g$).

This formula is approximately valid for collisions in which ions collide with slower moving electrons, which are the only collisions in which the ions lose energy. But for collisions of faster moving electrons with ions, where the electrons lose energy to the ions, the Coulomb logarithm, by the same logic, is Ln($mv_e^2/\hbar\omega_g$). If $v_e >> v_i$ then Ln($mv_e^2/\hbar\omega_g$) can be much larger than Ln($mv_i^2/\hbar\omega_g$) for sufficiently large values of $\hbar\omega_g$, in other words for sufficiently large B. Ignoring momentum transfer parallel to field, steady state occurs when $T_i/T_e$ = Ln($mv_e^2/\hbar\omega_g$)/Ln($mv_i^2/\hbar\omega_g$) [6].

This effect has been studied in a few cases for fusion plasmas with relatively weak fields, where is shown to be a relatively small effect [7]. It has been studied much more extensively in the case of neutron stars [8]. However, until the present research, it has not been applied to the DPF plasmoids, whose force-free configuration and very strong magnetic fields make the effect far more important. The author first demonstrated the importance of this effect for the DPF in 2003[1]. This paper reviews the earlier work and extends it with new simulations.

## 2. MAGNETIC FIELD EFFECT IN DPF PLASMOIDS

The dense plasma focus device produces hot-spots or plasmoids, which are micron-sized magnetically self-contained configurations with lifetimes of nanoseconds to tens of nanoseconds. It is within these plasmoids that the plasma is heated to high energy and fusion reactions take place. Such plasmoids have been observed to have magnetic field as high as 400 MG and density in excess of $10^{21}$/cc [1,9,10,11].

As shown in [1] and recapitulated here in Section 3, much higher densities and field strengths, into the multi-GG range, seem possible with the DPF.

To apply the magnetic effect to the DPF plasmoids, which are force-free configurations, we first note that small-angle momentum transfer parallel to the field can be neglected in these plasmoids, since the ion velocity lies very close to the local magnetic field direction, and $\Delta p_{par}/\Delta p_{perp} \sim \sin^2\theta$, where $\theta$ is the angle between the ion velocity and the B field direction[8].

In a force-free configuration, such as the toroidal vortices that make up the plasmoids, ions disturbed by collisions return to the local field lines in times of order $1/\omega_{gi}$, so

$$\theta \approx \omega_{ci}/\omega_{gi} \qquad (1)$$

Where $\omega_{ci}$ is the ion collision frequency. For a decaborane plasma, $\theta \sim 2\times 10^{-8} n/T_i^{3/2} B$. For the example of the plasmoid conditions obtained in [1], $n_i = 3\times 10^{21}$, B =400MG, $\theta$= 0.01 for $T_i$=60keV. For an example near break-even conditions, $n_i=1.4\times 10^{24}$, B =16 GG, $\theta$= 0.004 for $T_i$=600keV Small-angle parallel momentum transfer is significant only for combinations of very high $n_i$ and , $T_i$ < 60keV, which generally do not occur except during very brief early phases of the heating and compression of dense plasmoids, as we shall see in Sec. 4.

Even more significantly, the high B in plasmoids generates a regime where $mv_i^2/\hbar\omega_g<1$. In this case the magnetic effect is very large, the above formulae break down and quantum effects have to be considered. Such a situation has not been studied before for fusion applications, but has been analyzed extensively in the case of protons falling onto neutron stars[8].

In a strong magnetic field, since angular momentum is quantized in units of $\hbar$, electrons can have only discrete energy levels, termed Landau levels (ignoring motion parallel to the magnetic field):

$$E_b = (n+1/2)e\hbar B/mc = 11.6 eV(n+1/2)B(GG) \qquad (2)$$

Viewed another way, electrons cannot have gyroradii smaller than their DeBroglie wavelength. Since maximum momentum transfer is mv, where v is relative velocity, for $mv^2/2 < E_b$ almost no excitation of electrons to the next Landau level can occur, so very little energy can be transferred to the electrons in such collisions. Again ignoring the electron's own motion along the field lines, thus condition occurs when

$$E_i < (M/m)E_b \qquad (3)$$

For $E_i = 300\text{keV}$, this implies B>14GG for p, B>3.5GG for $\alpha$, and B>1.3GG for $^{11}$B. As will be shown below, such field strengths should be attainable with the DPF.

If we assume that $T_{eth} \gg E_b$, then we have to consider the motion of the electrons along the field lines, which can increase the relative velocity of collision, v. In the classical case, the ions will lose energy only from electrons for which $v_{epar} < v_i$. Since for these collisions $v < 2v_i$, energy loss will still be very small if $E_i < 1/2(M/m) E_b$, which can occur for boron nuclei.

However, there is a phenomenon which prevents energy loss to the electrons from falling to negligible levels. In the classical case, considering only motion along the line of force, an ion colliding with a faster moving electron will lose energy if the electrons' velocity is opposite to the of the ion, but will gain energy if they are in the same direction--the electron overtaking the ion. In the latter case the relative velocity is less than in the former case, and since the energy transfer increases with decreasing relative velocity, there is a net gain of energy to the ion. For an ion moving faster than the electron, the ion overtakes the electrons, and thus loses energy independently of the direction that the electron is moving in. Thus ions only lose energy to electrons moving more slowly than they are.

In the situation considered here, ions in some cases can lose energy to electrons that are moving faster than the ions. Consider the case of ions moving along the field lines colliding with electrons in the ground Landau level. If $v_{epar}$ is such that $m(v_i+v_{epar})^2 > 2 E_b$, while $m(v_i-v_{epar})^2 < 2E_b$, the energy lost by the ion in collision with opposite-directed electrons will much exceed that gained in same-directed collisions, since in the first case the electron can be excited to a higher Landau level, but not in the second case. In neither case can the electron give up to the ion energy from perpendicular motion, as it is in the ground state. (So, this consideration does not apply to above-ground-state electrons, which will lose energy to slower-moving ions.)

From these considerations, we can calcuatle an effectivce Couulomb logarithm, which is done in [1]. The result is present in Table 1

**Table 1**

| T | Ln$\Lambda$(T) |
|---|---|
| 0.05 | 0.536 |
| 0.1 | 0.525 |
| 0.2 | 0.508 |
| 0.3 | 0.491 |
| 0.4 | 0.48 |
| 0.5 | 0.473 |
| 0.6 | 0.47 |
| 0.8 | 0.474 |
| 1.0 | 0.488 |
| 2.0 | 0.631 |
| 3.0 | 0.803 |
| 4.0 | 0.964 |
| 6.0 | 1.244 |

For the heating of the ions by the much faster thermal electrons, with $T_e \gg 1$, quantum effects can be ignored and the coulomb logarithm is simply $Ln(2T_e)$.

## 3. CONDITIONS IN DPF PLASMOIDS

To see what the consequences of the magnetic field effect are for DPF functioning, we first use a theoretical model of DPF functioning that can predict conditions in the plasmoid, given initial conditions of the device. As described by Lerner [12], and Lerner and Peratt [13], the DPF process can be described quantitatively using only a few basic assumptions. Using the formulae derived there, Lerner [1] showed that the particle density increases with $\mu$ and z as well as with I, and decreases with increasing r. Physically this is a direct result of the greater compression ratio that occurs with heavier gases, as is clear from the above relations. Thus the crucial plasma parameter $n\tau$ improves with heavier gases.

The theoretical predictions of the formulae in [1] are in good agreement with the results obtained experimentally ,cited in the same paper. If we use these equations to predict $B_c$ the magnetic field in the plasmoid, we obtain 0.43 GG, in excellent agreement with the observed value of 0.4 GG. Similarly, the formulae yield $n\tau = 4.6 \times 10^{13}$ sec/cm$^3$ as compared with the best observed value of $9 \times 10^{13}$ and the average of $0.9 \times 10^{13}$.

For decaborane with z=2.66 and u =5.166, with r= 5 cm, I =3MA, the formulae in [1] yield B =12GG and nt =$6 \times 10^{15}$. This is of course a considerable extrapolation-- a factor of 60 above the observed values in both B and $n\tau$. However, these conditions can be reached with relatively small plasma focus devices.

The limit on the achievable magnetic field is set mainly by the mechanical strength of the electrodes. Since $B_c$ for a given fill gas is proportional to $B_i$, the field at that cathode, a small cathode radius is desirable. For proper DPF functioning, the anode radius generally must be no more that 0.3-0.5 times the cathode radius. However, the anode is subject to thermal-mechanical stress due to the transient heating and expansion of the outer layer by the discharge current.

Thermal-mechanical simulations of the anode show that elastic stress limits will be reached for copper electrodes at anode field strengths above 200kG for cooper electrodes and 380 kG for beryllium electrodes. With a cathode/anode radius ratio fo 2.5, this implies a maximum $B_c$ of 15 GG with a beryllium anode.

## 4. SIMULATION OF PLASMOIDS

To determine if net energy production is feasible with pB11, a zero-dimensional simulation has been run that contains the relevant physics, including the magnetic field effect. While not fully realistic, the simulation is adequate to show the impact of the magnetic effect and the possibility for high fusion yields.

The simulation, by its zero-dimensional character, assumes that the plasma in the plasmoid is homogenous. In addition the simulation assumes Maxwellian distributions for the electrons, and hydrogen and boron ions. Helium ions, produced by the fusion reaction, are assumed to cool to a Maxwellian distribution, but the fusion alpha particles are treated separately, as described below, as they are slowed by the plasma. In accordance with observation, it is assumed that the ions are all fully ionized.

Initial conditions include the radius and magnetic field strength of the plasmoid, the initial density and average energy (or temperature) of electrons, hydrogen and boron ions (which are summed to be pure B11). The model assumes a smooth, sinusoidal increase in magnetic field and a corresponding decrease in radius. After the magnetic field peaks, the radius is assumed to be constant. Stable confinement is also assumed, again in accord with experimental results. At each time step, electron and ion beams are generated, which evacuate particles from the plasmoid and subtract energy from the magnetic field. Following the formulae above, at each instant the current in the beam is $I_c/4\pi^2$ and beam power is $5.5\mu^{-3/4}I_c/4\pi^2$ W. Beam power is subtracted from the total magnetic field energy at each time step. All the energy in the electron beam however, is assumed to heat the plasma electrons, so does not leave the plasmoid. This assumption is justified by experimental results at much lower plasmoid densities than modeled here, which indicate that nearly all electron beam energy is transferred to the plasmoid electrons [1].

For each time step, the model calculates the x-rays emitted by the electron and the energy exchange between the ions and electron and between the ions species, with each species having its own temperature. The magnetic effect is included using an approximation to the numerical values calculated above for the coulomb logarithm term. The approximation used is

$$\ln \lambda = -0.0561T^3 + 0.268T^2 - 0.2729T + 0.5507 \; for \; T < 1.3$$
$$\ln \lambda = 0.02(\ln T)^3 + 0.265(\ln T)^2 + 0.0045\ln T + 0.5045 \; for \; T > 1.3$$

(4)

Where T is the dimensionless ion temperature, as defined above. This formula is valid up to T=100, well beyond what occurs in these runs.

During the early part of the simulation run, when ion temperature is still relatively low, <60keV, the average angle between ion velocity and magnetic field, $\theta \sim 2 \times 10^{-8} n/T_i^{3/2}B$, is significant, leading to a much larger $\lambda$. However, ion temperature rises so rapidly that $\theta$ becomes insignificant after ~0.4 ns, out of multi-ns runs. Detailed comparisons indicate that this factor has negligible effect on the simulation outcomes, so it has not been included.

Using the calculated temperatures and densities, the model calculates the thermonuclear reaction rate, based on published rates and cross sections (Cox, 1991). It then calculates the transfer of energy by individual collisions to the ions and electrons, using the Coulomb logarithm of $Ln(2T_e)$ for the electrons.

As the fusion reactions increase the thermal energy of the ions and electrons, $\beta$, the ratio of magnetic field energy to thermal energy decreases. When $\beta=1$, the model assumes that the plasmoid radius remains constant and half the thermal power is converted into additional magnetic field energy to maintain a constant radius. This is of course a gross approximation, since we can not determine with a zero-dimension simulation whether the plasmoids will in fact remain stable, expand slowly, or completely disrupt as the thermal energy increases.

A number of simulations were run, with a wide range of parameters. The result for a "nominal" run are first described, then an analysis of how these results change with various parameters and by modifying the model assumptions. The nominal simulation has the parameters described in Table 2

**Table 2 Nominal Simulation Parameters**

| Peak Plasmoid magnetic field | 13 GG |
| Plasmoid core radius | 8.63 microns |
| Peak Electron Density | $3.8 \times 10^{24}$/cc |
| Time to peak compression | 0.7 ns |
| Fuel | $B_{10}H_1$ |

These parameters were selected to produce a high fusion yield, yet be achievable with existing large DPF devices.

As seen in Figure 1, the ratio of ion to electron temperature rises rapidly to a peak of around 17, quite close to the figure anticipated by the ratios of the coulomb logarithm in the above magnetic field effect calculations. As a result, by t=0.9 ns, thermonuclear power exceeds x-ray power (at a point where the proton/electron energy ratio already exceeds 11) and rapid heating of the ions occurs . The thermonuclear /x-ray power ratio peaks at 2.2 at t= 1.4 ns.

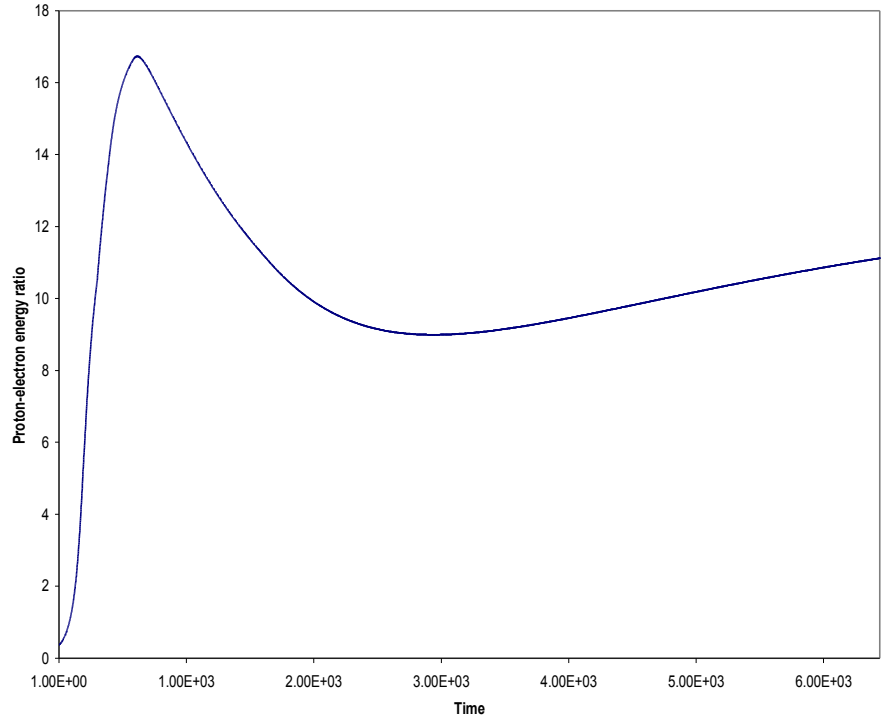

**Figure.1**. Ratio of ion to electron average energy vs time. Time is in steps of 30 ps.

As the magnetic field rises, the electron temperature actually momentarily falls, but then rises again and reaches a peak of 57 keV at 2.0 ns, while the ion temperature passes right through the optimal temperature of 600keV and peaks at over 940keV. Thermonuclear power has already peaked at 1.4 ns as the ion temperature passes through the optimal 600keV level. Ion and electron temperature start to decrease as thermal energy in the plasmoid matches magnetic confinement energy and the particles start to do work against the magnetic field.

Most of the thermonuclear burn occurs as the plasmoid cools down and feeds its energy into the ion beam and x-rays, so that at the peak of thermonuclear power production only 20% of the fuel in the plasma has been fused, but by the end of significant thermonuclear burn at around 18 ns, 82.5% of the fuel has been burned. The x-ray and beam pulses decay nearly exponentially, but are at less than 1% of peak power by 15 ns and have a FWHM of 3 ns.

Total input energy in this example is 14.6 kJ, x-ray yield is 9.5 kJ and beam yield is 13.4 kJ, so total output energy exceeds input energy by a ratio of 1.57.  Preliminary estimates indicate that energy conversion of both the x-rays and the ion beam can reach 80% with proper design, so that net energy production with close to 50% thermodynamic efficiency should be possible, if other losses in the entire system can be reduced to levels small in comparison.

Leaving the model assumptions the same, the magnetic field of the plasmoid can be varied. This is the equivalent of varying the peak current of the DPF for fixed electrode dimensions. The results are presented in Table 3. As shown there, the total output/input ratio continues to rise slowly above 13 GG but the beam output starts to decline above 11 GG, due to the increasing density of the plasma and thus increasing x-ray emission relative to beam power. If, as is likely, energy conversion of the ion beam is more efficient than conversion of the x-ray pulses, 13 GG may be more desirable than higher fields, which may also be more difficult to achieve.

**Tabl e 3**

| Peak I (MA) | 1.9 | 2.3 | 2.8 | 3.2 | 4.3 |
|---|---|---|---|---|---|
| B (GG) | 9.0 | 11.0 | 13.0 | 15.0 | 20.0 |
| Gross Input (kJ) | 7.0 | 10.5 | 14.6 | 19.5 | 34.6 |
| X-ray/Input | 0.49 | 0.55 | 0.65 | 0.75 | 1.00 |
| Beam/Input | 0.88 | 0.93 | 0.92 | 0.87 | 0.68 |
| Beam+x-ray/Input | 1.37 | 1.48 | 1.57 | 1.62 | 1.68 |

Larger changes in yield are obtained when some of the assumption of the model is modified. The density of the plasmoid is set, as described in section 3, by the condition that the electron gyrofrequency is twice the plasma frequency, so the synchrotron radiation is just barely trapped. The nominal cases calculate the density based on the condition being met in the case of low-energy electrons. However, since the synchrotron frequency increases as $\gamma^2$, where $\gamma$ is the relativistic factor and the plasma frequency decreases as $\gamma^{-0.5}$, the ratio of the two frequencies increases as $\gamma^{2.5}$ and the density for a given B increases as $\gamma^5$.

For example for $T_e$ of 30 keV, critical trapping for the synchrotron radiation occurs at a density 1.32 times that of the nominal cases. In this case, x-ray yield increase to 1.05x input, beam is 0.86x input and I/O ratio increases to 1.90, an increase of 20%. It should be noted, however that there are not experimental results that indicate the higher $n/B^2$ ratio is actually obtained and they are instead consistent with the nominal ratio. This is probably due to the fact that, in the force-free plasmoid configuration, the electrons travel along the lines of force, whose curvature is considerably greater than the gyro-radius.

On the other hand, results may be decreased if we modify the assumption that the plasmoid will not expand once thermal energy become equal to magic energy. If we instead assume that current in the plasmoid will neither increase nor decrease due to the increased thermal energy, the plasmoid will gradually increase in radius, decreasing density and quenching the fusion reaction.

In this case, x-ray yield declines to 0.44x input while beam yield is almost unchanged at 0.88x input and total yield declines to 1.32, 17% less than the nominal case and too low, probably, for net energy production with reasonable conversion efficiencies. However, there is good reason to believe that this constant-current case is unduly pessimistic, since the decrease in B due to expansion would induce electric fields that will increase the current, resisting expansion.

## 5. DRIFT KINETIC FLUID PARTICLE SIMULATION

The preliminary zero-dimensional simulation has significant limitations that must be overcome to achieve accurate predictions of DPF functioning. First, since we want to maximize the transfer of energy into the plasmoid, we need a simulation that models the whole DPF process, including the run-down phase and the formation of the plasmoid. Second, we need to understand how the plasmoid will actually react to the release of dynamically significant amounts of fusion energy. For both these purposes we need 3-D simulations.

However, the DPF presents real challenges for simulators that prevent conventional techniques from being effective. The DPF involves processes that extend over a large range of time and space scales, from the sub-ns evolution times and micron spatial scales of the plasmoids to the microsecond time scale and multi-cm spatial scales of the rundown and collapse phases. The plasma is non-collisional, with gyroradius being smaller than mean-free-path, which means that MHD approximations do not give correct answers, and cannot simulate the main features of filament formation, pinch collapse and plasmoid formation and decay. But particle-in-cell simulations are not only handicapped by the wide range of scales, but even more by the need to resolve frequencies as high as the electron gyrofrequency, which can be as high as $10^{17}$Hz.

To overcome the limitations of conventional approaches, our novel computational approach to a more robust and physically accurate model than that explored above rests with a new drift kinetic fluid particle (DKFP) method, grounded in what are essentially exact solutions in the limit of Knudsen or Vlasov flows.

The following discussion shows how drift kinetic fluid particles are built up from fundamental Vlasov solutions and connected to observables from other parts of a model or from experimental data. First, we set a local drift velocity and its gradient, together with local acceleration and its gradient,

$$\mathbf{C} = \mathbf{V} + \mathbf{x} \cdot \delta/h \,, \mathbf{a} = \mathbf{A} + \mathbf{x} \cdot \alpha/h \,, \tag{5}$$

as functions of local position within a box. The velocity field is just the familiar mean molecular velocity, and the acceleration field is just the local imposed force on the particle species under consideration — electromagnetic, gravitational, or whatever. Next, the single particle distribution functions are built up as outer products on the cartesian components of configuration and velocity space,

$$f(\mathbf{x}, \mathbf{v}) = \Pi_i^3 \mathcal{B}_{wi}(X, x) \mathcal{G}_{ui}(V, v) \tag{6}$$

where the box in space extends initially over a width h in each dimension and the velocity function has a characteristic thermal speed U, perhaps distinct in each velocity dimension. As written the distribution function in (6) solves the Vlasov operator equation

$$(\partial_t + \mathbf{V} \cdot \partial_{\mathbf{X}} + \mathbf{A} \cdot \partial_{\mathbf{V}})f \equiv \mathcal{L}_0 f = 0 , \tag{7}$$

exactly, in terms of the characteristics

$$\mathbf{X} = \mathbf{x} + \int^t dt_1 \mathbf{v} + \int^{t_1} dt_2 \mathbf{a}$$

And

$$\mathbf{V} = \mathbf{v} + \int^t dt_1 \mathbf{a} .$$

Now, any velocity moment can be written as

$$\mathcal{O}_L(X, t) = \left(\frac{N}{2h}\right)\frac{1}{U\sqrt{2\pi}} \int_{-h}^{h} \left(\frac{dx}{t}\right) \left(\frac{X-x}{t}\right)^L \times \exp\left(-\frac{1}{2}\left(\frac{(((X-(A+\alpha x/h)t^2/2-x)/t)-(V+\delta x/h))}{U}\right)^2\right) \tag{8}$$

And, with the auxiliary variables,

$$D(t) = \left(1 + \frac{t\delta}{h} + \frac{1}{2}\frac{t^2 \alpha}{h}\right) ,$$
$$h_+(t) = h + (V + \delta)t + \frac{1}{2}(A + \alpha)t^2 ,$$
$$h_-(t) = -h + (V - \delta)t + \frac{1}{2}(A - \alpha)t^2 ,$$

and one finds a density function

$$n(X, t) = \frac{N}{2hD(t)} \left(\text{erf}\left(\frac{(h_+(t) - X)}{\sqrt{2}U t}\right) + \text{erf}\left(\frac{(X - h_-(t))}{\sqrt{2}U t}\right)\right) . \tag{9}$$

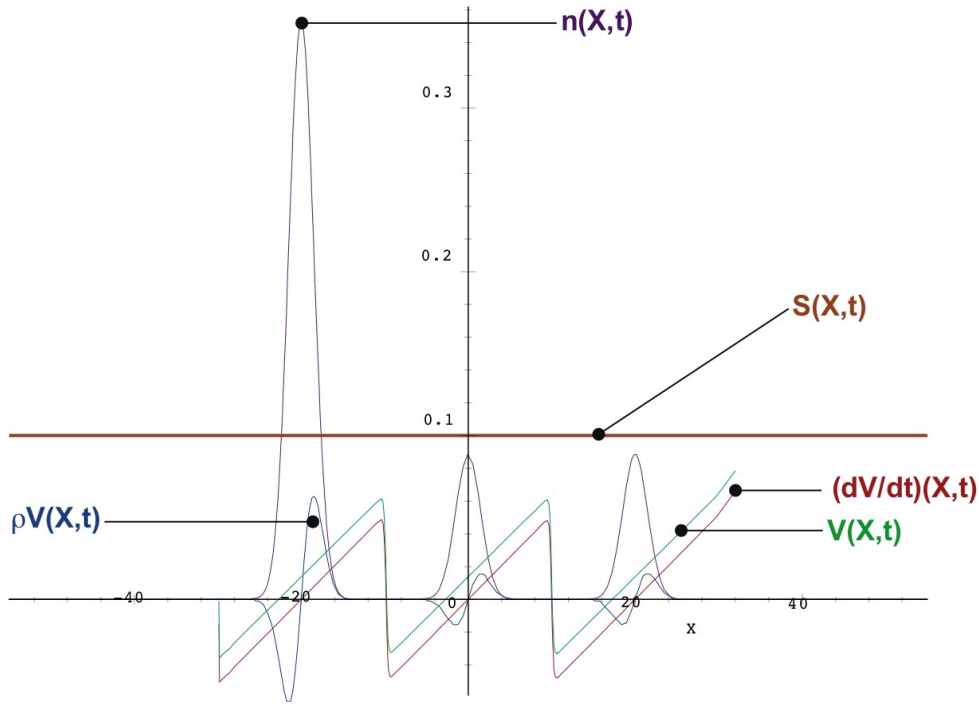

**Figure.2**. Three fluid particles collide. The acceleration and velocity fields are everywhere proportional and the similarity S(x,t) is everywhere constant.

The expected profiles of fluid variables and the usual moments are now all now available in closed form on all space and time as a particle set evolves. The action of pressure gradients is built in as these particle interpenetrate, capturing the fluid pressure automatically as a superposition of partial pressures, as shown in Fig.2.

Collisions are introduced in either of two ways. First, in a weakly collisional regime, the interactions may be described in relaxation time approximation. The momentum transfer, elastic scattering, or even charge exchange cross sections are transformed into collision frequencies and imposed on the drifting particles as apparent drag forces. This method can be viewed as a sort of partial operator spitting insofar as those collision frequencies can be updated without ever reforming the particle population.

Alternately, in a strongly collisional regime, transport fluxes are computed into a target volume and the implied time variations on any thermodynamic variables of interest are computed for that volume from the collision operators required by the chosen formulation. Here a complete operator split is demanded insofar are the particle population must be annihilated and reformed with new values for densities, mean velocities and thermal speeds.

The strength of this method in relation to the problem of rapidly evolving plasmoids is in the great freedom available to treat a wide range of fluid regimes concurrently on a wide range of spatial scales—with the same basic formulation. We intend to develop models based on this method to initially simulate the formation of plasmoids.

## ACKNOWLEDGEMENTS

Work performed for the U. S. Department of Energy under Contract No. W--7405--ENG--36.